\documentclass[twocolumn,prl,tightenlines,superscriptaddress,reprint]{revtex4-2}
\usepackage{hyperref}
\usepackage{amsmath,amssymb}
\usepackage[utf8]{inputenc} 			
\usepackage[T1]{fontenc}				
\usepackage{graphicx}
\usepackage{float}
\usepackage[usenames,dvipsnames]{xcolor}
\usepackage{tikz}
\usepackage{ulem}

\newcommand{\eps}{\varepsilon}

\newcommand{\grad}{\nabla}
\renewcommand{\div}{\nabla\cdot}
\renewcommand{\vec}[1]{\mathbf{#1}}
\newcommand{\Gvec}[1]{\boldsymbol{#1}}

\newcommand*\colvec[3][]{
    \begin{pmatrix}\ifx\relax#1\relax\else#1\\\fi#2\\#3\end{pmatrix}
}

\newcommand{\nperp}{\vec n_\bot}
\newcommand{\dpar}{\partial_\parallel}
\newcommand{\dperp}{\partial_\bot}
\newcommand{\qpar}{q_\parallel}
\newcommand{\qperp}{q_\bot}

\usepackage{pdfpages}
\usepackage{pgffor}
\makeatletter
\AtBeginDocument{\let\LS@rot\@undefined}
\makeatother

\begin{document}

\title{Dynamic Scaling of Two-Dimensional Polar Flocks}

\author{Hugues Chat\'{e}}
\affiliation{Service de Physique de l'Etat Condens\'e, CEA, CNRS Universit\'e Paris-Saclay, CEA-Saclay, 91191 Gif-sur-Yvette, France}
\affiliation{Computational Science Research Center, Beijing 100094, China}
\affiliation{Sorbonne Universit\'e, CNRS, Laboratoire de Physique Th\'eorique de la Mati\`ere Condens\'ee, 75005 Paris, France}

\author{Alexandre Solon}
\affiliation{Sorbonne Universit\'e, CNRS, Laboratoire de Physique Th\'eorique de la Mati\`ere Condens\'ee, 75005 Paris, France}

\date{\today}
\begin{abstract}
  We propose a hydrodynamic description of the homogeneous ordered
  phase of polar flocks.  Starting from symmetry principles, we
  construct the appropriate equation for the dynamics of the Goldstone
  mode associated with the broken rotational symmetry. We then focus
  on the two-dimensional case considering both ``Malthusian flocks''
  for which the density field is a fast variable that does not enter
  the hydrodynamic description and ``Vicsek flocks'' for which it
  does. In both cases, we argue in favor of scaling relations that
  allow to compute exactly the scaling exponents, which are found in
  excellent agreement with previous simulations of the Vicsek model
  and with the numerical integration of our hydrodynamic equations.
\end{abstract}

\maketitle

A key result of active matter studies is the possible emergence, in two dimensions, 
of long-range polar order among locally aligning self-propelled particles~\cite{vicsek_novel_1995,toner_long-range_1995,chate_collective_2008,mahault_quantitative_2019,chate_dry_2020},
something unattainable in equilibrium, as proven by the
Mermin-Wagner-Hohenberg
theorem~\cite{mermin_absence_1966,hohenberg_existence_1967}.  However,
the properties of the resulting symmetry-broken phase, in particular
the temporal and spatial correlations of fluctuations, have remained
elusive. In their seminal 1995 paper~\cite{toner_long-range_1995},
Toner and Tu (TT) argued that the scaling exponents can be computed
exactly in $d=2$ dimensions, but this argument was later found to fail
due to terms that had been missing in the original
theory~\cite{toner_reanalysis_2012}.  There was nevertheless hope that
the TT95 exponents would still be correct if these new terms were to
be renormalized so as to become
irrelevant~\cite{toner_reanalysis_2012}. However, large-scale
simulations of the Vicsek model clearly displayed a different
scaling~\cite{mahault_quantitative_2019}, calling for a refined
theory.

The TT95 exponents were nevertheless thereafter believed to control
the scaling of fluctuations of $d=2$ ``Malthusian flocks'' within
which particles reproduce and die on a short time
scale~\cite{toner_birth_2012}, so that, on large scale, the density is
effectively constant and drops out of the hydrodynamic description.
In a recent paper where we showed that the ordered phase of Malthusian
flocks, in the limit of strictly constant density, is in fact
metastable~\cite{besse_metastability_2022}, we also found exponent
values compatible with the TT 1995 predictions.

So far, to our best knowledge, all studies of the ordered phase have
started from the ``isotropic'' Toner-Tu equations (describing the
system in all phases) expanded around the ordered state, eliminating
the fast degrees of
freedom~\cite{toner_long-range_1995,toner_flocks_1998,toner_birth_2012,toner_reanalysis_2012,chen_universality_2020,chen_moving_2020}.

In this Letter, we take a different route, writing directly a generic
equation for the dynamics of the hydrodynamic modes of the ordered
phase. We first construct the equation for the $d-1$ Goldstone modes
associated with the spontaneously broken continuous symmetry in
arbitrary dimension $d$.  We then focus on the $d=2$ case, considering
both the density-less (Malthusian) case and the case relevant to the
Vicsek model, when the Goldstone mode is coupled to a density
field. Our approach highlights the symmetries of these equations,
which ---somewhat surprisingly--- were not apparent in previous works.
We obtain simpler equations that, importantly, possess a different
structure which allows us to derive the scaling exponents in $d=2$ for
both universality classes.  Moreover, the numerical integration of our
hydrodynamic theory provides a more efficient way to measure the
fluctuations of the ordered phase than when using the isotropic
equation. We take advantage of this to show that the scaling exponents
that we derived agree well with both numerical measurements from our
hydrodynamic theory and the large-scale simulations of the Vicsek
model of Ref.~\cite{mahault_quantitative_2019}.

We first consider the simpler case of the Malthusian flocks introduced
by Toner~\cite{toner_birth_2012}. Self-propelled particles align
locally, but also reproduce and die. As a result of this population
dynamics, the density field is a fast mode and thus drops out of the
hydrodynamic description, which then only involves a velocity field
$\vec v$ that does {\it not} obey the incompressibility condition
$\div \vec v=0$. On the contrary, it can be seen as infinitely
compressible~\cite{di_carlo_evidence_2022} and its dynamics are akin
to Burgers' equation with additional alignment terms. In the simple
form considered in Ref.~\cite{besse_metastability_2022} that contains
the essential terms it reads
\begin{equation}
  \label{eq:full-eq}
  \partial_t \vec v+\lambda(\vec v\cdot \nabla)\vec v=D\nabla^2\vec v+(r-u|\vec v|^2)\vec v +\sqrt{2\Delta}\,\Gvec\xi \,,
\end{equation}
where $\Gvec\xi$ is a Gaussian white noise field with correlations
\begin{equation}
  \label{eq:corre-noise}
  \langle \xi_\alpha(\vec r,t) \xi_\beta(\vec r',t') \rangle =\delta^d(\vec r-\vec r') \delta(t-t') \delta_{\alpha\beta}.
\end{equation}
with Greek indices denoting Cartesian coordinates. At large enough
alignment strength $r$, one can observe a long-range-ordered phase
spontaneously breaking the rotational symmetry. Although this phase
was found in ~\cite{besse_metastability_2022} to be metastable to the
nucleation of topological defects, such nucleation, in a large part of
the phase diagram, happens very rarely. The scale-free anisotropic
correlations of the phase described in~\cite{toner_birth_2012} can
then be observed at will.

The hydrodynamic variable that emerges from the spontaneous symmetry
breaking is a direction in space which we represent by the unit vector
field $\vec n(\vec r,t)$. 
Let us construct a dynamical equation for
$\vec n(\vec r,t)$ that contains all terms allowed by symmetry. As a
unit vector, it can only rotate so the dynamics should read
\begin{equation}
  \label{eq:rotation}
  \partial_t \vec n=\vec R\times\vec n,
\end{equation}
with rotation vector $\vec R$. Because the symmetry is spontaneously
broken, the deterministic part of $\vec R$ can only come from
differences with the local environment. Mathematically, this means that
it must be expressed as the divergence of a rank-2 tensor
${\mathcal R}$. Adding a noise term, the rotation vector then reads
\begin{equation}
  \label{eq:rotation-div}
  R_\alpha=\partial_{\beta}({\mathcal R}_{\alpha\beta})+\sqrt{2\Delta}\xi_\alpha(\vec r,t)
\end{equation}
with $\xi_\alpha$ a Gaussian white noise with correlations given by
Eq.~(\ref{eq:corre-noise}). We then write all possible terms for
$\mathcal R$ at first order in gradients that respect the $O(d)$
rotational symmetry. They all need to involve the Levi-Civita tensor
$\eps_{\alpha\beta\gamma}$ for $\vec R$ to be an axial vector. We
obtain
\begin{align}
  \label{eq:all-terms-R}
  &{\mathcal R}_{\alpha\beta}=-\lambda \eps_{\alpha\beta \gamma}n_\gamma \nonumber\\
  &+\eps_{\alpha \gamma \delta}\left[-D_1 (\partial_\beta n_\gamma)n_\delta-D_2 (\partial_\gamma n_\beta)n_\delta+D_3 (\partial_\gamma n_\delta)n_\beta\right] \nonumber\\
  &+\eps_{\beta \gamma \delta}\left[ D_4 (\partial_\gamma n_\delta)n_\alpha+D_5(\partial_\alpha n_\gamma)n_\delta\right] \nonumber\\&+\eps_{\gamma\alpha\beta}\left[ D_6 (\partial_\delta n_\gamma)n_\delta-D_7(\partial_\delta n_\delta)n_\gamma\right]. 
\end{align}
Note that for a passive system $\vec n$ is not
coupled to motion in space and Eq.~(\ref{eq:all-terms-R}) would then
need to obey separately the spatial $O(d)$ symmetry and the $O(n)$
symmetry of $\vec n$ (with possibly $n\neq d$). This would preclude all
terms except $D_1$, the only one in which there is no contraction
between a spatial derivative and the vector $\vec n$.

We now consider specifically the case $d=2$, parameterizing
$\vec n=(\cos\theta,\sin\theta,0)$. Eq.~(\ref{eq:rotation}) becomes
$\partial_t\vec n=\partial_t \theta \nperp= R_z \nperp$ with
$\nperp=(-\sin\theta,\cos\theta,0)$. After computing $R_z$ from
Eq.~(\ref{eq:all-terms-R}), it gives
\begin{align}
  \partial_t \theta = &\div\left[ \lambda \vec \nperp  +D_1\nabla\theta +(D_3 +D_6)\dpar\theta \vec n \right. \nonumber \\ & \quad \left. +(D_2+D_7)\dperp\theta \vec \nperp \right] +\sqrt{2\Delta}\xi_z
  \label{eq:theta2d}
\end{align}
where $\dpar\equiv \vec n\cdot \grad$ and
$\dperp\equiv \nperp\cdot \grad$. Interestingly, we show
in~\cite{Supp} that, adiabatically eliminating the norm of $\vec v$
(the fast variable) from Eq.~(\ref{eq:full-eq}), the resulting
equation for its phase $\phi$ is not exactly Eq.~(\ref{eq:theta2d})
because it contains diffusion terms which cannot be written as a
divergence. However, the dynamics of $\tilde\phi=\phi-\alpha\dpar\phi$
for a well-chosen value of $\alpha$ does take the form of
Eq.~(\ref{eq:theta2d}). This points to the surprising fact that the
Goldstone mode is {\it not} the phase $\phi$ of the velocity, as one
would naively expect, but, rather, the combination $\tilde \phi$.

In the following, we neglect for simplicity the terms with
coefficients $D_2$, $D_3$, $D_6$ and $D_7$.  (They lead to anisotropic
diffusion terms, which we write explicitly in~\cite{Supp}, that are
not expected to qualitatively change the large-scale behavior, as
will become clear below.)  Denoting $D_1\equiv D$ and
$\xi_z\equiv \xi$, Eq.~(\ref{eq:theta2d}) then rewrites as
\begin{equation}
  \label{eq:theta2d-finale}
  \partial_t \theta +\lambda \dpar\theta= D\nabla^2\theta + \sqrt{2\Delta}\xi
\end{equation}
which we believe to be the simplest equation capturing the universal
features of Malthusian flocks.

Eq.~(\ref{eq:theta2d-finale}) still possesses $O(2)$ rotational
symmetry with $\theta$ taking arbitrary values.  To investigate the
scaling behavior of fluctuations around the ordered state, we assume
that the system is globally ordered along the $x$-axis (corresponding
to $\theta=0$). Shifting to the comoving frame
$\vec r\to \vec r-\lambda \vec {e_x}$ and expanding the nonlinearities
contained in the convective derivative in
Eq.~(\ref{eq:theta2d-finale}) yields, at order $\theta^3$:
\begin{equation}
  \label{eq:theta2d-finale-exp}
  \partial_t \theta+\lambda_y\partial_y \theta^2+\lambda_x\partial_x\theta^3 = D_x\partial_x^2\theta+ D_y\partial_y^2\theta + \sqrt{2\Delta}\xi
\end{equation}
where we have introduced generic coefficients taking the bare values
$\lambda_y=\lambda/2$, $\lambda_x=-\lambda/6$ and $D_x=D_y=D$. An
important difference between Eq.~(\ref{eq:theta2d-finale-exp}) and the
one obtained in Ref.\cite{toner_birth_2012} is the presence here of
the $\lambda_x$ term which we believe was unduly neglected before,
being both allowed by symmetry and relevant in the renormalization
group (RG) sense.  Upon performing a RG step, integrating the
short-distance fluctuations over an infinitesimal momentum shell
$\Lambda/b\le |\vec q| \le \Lambda$ with $b=1+ds$, $\Lambda$ the
ultraviolet cutoff, and rescaling
\begin{equation}
  \label{eq:rescaling}
 y\to by,\,\, x\to b^\zeta x,\,\, t\to b^z t,\,\, \theta\to b^\chi\theta,
\end{equation}
the coefficients of \eqref{eq:theta2d-finale-exp} evolve under the
RG flow equations
\begin{align}
    &\frac{d\lambda_x}{ds}=[2\chi \!+\! z \!-\! \zeta \!+\! \eta_{\lambda_x}]\lambda_x;\,\, 
    \frac{d\lambda_y}{ds}=[\chi \!+\!z \!-\! 1 \!+\! \eta_{\lambda_y}]\lambda_y; \nonumber\\
  & \frac{d D_x}{ds}=[z \!-\! 2\zeta \!+\! \eta_{D_x}] D_x; \;\;\;\;
  \frac{dD_y}{ds}=[z \!-\! 2 \!+\! \eta_{D_y}] D_y; \nonumber \\
  &\frac{d\Delta}{ds}=\tfrac{1}{2}[z \!-\! 2\chi \!-\! 1 \!-\! \zeta +2\eta_{\Delta}] \Delta 
  \label{eq:RG-flow}
\end{align}
with the anomalous dimensions $\eta_{\ldots}$ coming from
``graphical'' corrections. Remarkably, the scaling exponents can be
determined exactly. First, since the nonlinearities of
\eqref{eq:theta2d-finale-exp} are derivatives, the noise term does not
get graphical corrections, and thus $\eta_\Delta=0$. Moreover, as we
discuss in~\cite{Supp}, we believe that the terms coming from the
convective derivative also do not receive graphical corrections so
that $\eta_{\lambda_x}=\eta_{\lambda_y}=0$ because of a generalized
Galilean invariance. This implies three scaling relations at the
infrared fixed point
\begin{equation}
  \label{eq:scaling-relations}
  2\chi+z-\zeta=\chi+z-1=z-2\chi-1-\zeta=0
\end{equation}
from which one obtains the values of the scaling exponents summarized
in Table~\ref{tab:exponents-malth}. Importantly, the field scales with
a negative exponent $\chi$, confirming that the system possesses
long-ranged order and that higher-order nonlinearities are
irrelevant.

\begin{table}[t!]
	\centering
	\caption{Scaling exponents for $d=2$ Malthusian flocks
          obtained from our prediction
          Eq.~(\ref{eq:scaling-relations}), from the prediction of
          Toner~\cite{toner_birth_2012} and from the correlation
          functions shown in
          Fig.~\ref{fig:corre-cst-density}(b-d). The fitting procedure
          leading our estimates is detailed in~\cite{Supp}.}
  \begin{ruledtabular}
  \begin{tabular}{cccc}
   \hspace{1cm} & This work & Toner 2012~\cite{toner_birth_2012}  & Numerics \\
   \hline
  $\chi$ & \!\!\!\!\!$-1/4$ & \!\!\!\!\!$-1/5$ & \!\!\!\!\!$-0.25(1)$\\
  $\zeta$ & $3/4$ &  $3/5$ & $0.75(2)$\\
  $z$ & $5/4$ & $6/5$ & $1.27(3)$\\
\end{tabular}
\end{ruledtabular}
 \label{tab:exponents-malth}
\end{table}

Let us now compare our findings to numerical simulations. To extract
the scaling behavior, we compute the Fourier spectrum of equal time
correlations $\langle |\theta(\vec q,t)|^2\rangle$ which shows the
anisotropic scaling~\cite{toner_flocks_1998}
\begin{equation}
  \label{eq:ani-scaling}
  \langle |\theta(\vec q,t)|^2\rangle \underset{q\to 0}{\sim}
  \begin{cases}
    \qpar^{-(1+2\chi+\zeta)/\zeta} \;\; \text{for}\; \qpar \gg \qperp^{\zeta}\\
    \qperp^{-(1+2\chi+\zeta)} \;\;\;\;\; \text{for}\; \qperp^{\zeta}\gg \qpar \;.
  \end{cases}
\end{equation}

Numerical simulations of \eqref{eq:theta2d-finale} or
\eqref{eq:theta2d-finale-exp} are much simpler than those of the
isotropic Eq.~(\ref{eq:full-eq}).  Indeed, a practical problem in
measuring the correlations \eqref{eq:ani-scaling} is that the
direction of global order slowly diffuses over time.  Strategies to
cope with this, discussed in detail
in~\cite{mahault_quantitative_2019}, include applying a large enough
external field to pin the order along one
direction~\cite{besse_metastability_2022} or, for particle systems,
considering closed boundary conditions in one spatial
direction~\cite{tu_sound_1998,kyriakopoulos_leading_2016,mahault_quantitative_2019}.
All of them affect the dynamics and reduce significantly the range
over which unspoiled scaling is observed. Moreover, these methods are
imperfect since the direction of order is not strictly pinned and
still fluctuates. On the contrary, imposing
$\int d\vec r \xi(\vec r,t)=0$ (in practice setting $\xi(\vec q=0)=0$
in Fourier space) at all times when integrating
\eqref{eq:theta2d-finale} or \eqref{eq:theta2d-finale-exp} cancels
global rotations and leaves the dynamics of interest intact.

\begin{figure}[t!]
  \centering
    \includegraphics[width=\columnwidth]{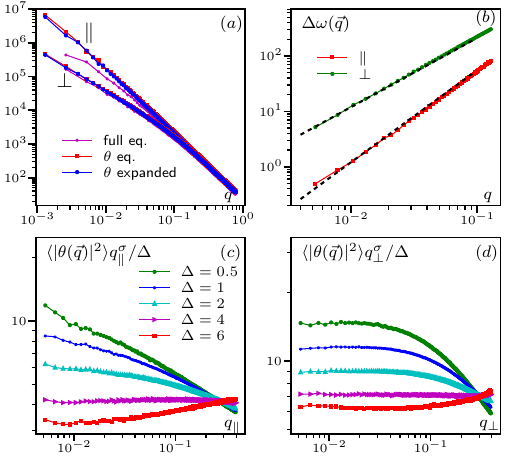}
 \caption{Correlations of fluctuations in $d=2$ Malthusian flocks.
   (a) Static correlation function \eqref{eq:ani-scaling} in the
   longitudinal direction $\vec q=(q,0)$ and perpendicular direction
   $\vec q=(0,q)$ measured in the isotropic Eq.~(\ref{eq:full-eq}),
   the hydrodynamic equation for $\theta$
   Eq.~(\ref{eq:theta2d-finale}) and its expanded version at order
   $\theta^3$, Eq.~(\ref{eq:theta2d-finale-exp}).  System size
   $L=2400$, $r=u=1$ for the full equation, $L=4800$ for the
   others. $\Delta=0.5$, $\lambda=D=1$ for all. (b) Width of the
   peaks in the $\omega$ spectra of
   $\langle |\theta(\vec q,\omega)|^2\rangle$ measured in
   Eq.~(\ref{eq:theta2d-finale-exp}) ($L=2400$, $\Delta=4$, dashed
   lines are our theoretical predictions).  (c-d) Static correlation
   functions measured in simulations of
   Eq.~(\ref{eq:theta2d-finale-exp}), rescaled by the predicted
   exponent $\sigma=(1+2\chi+\zeta)/\zeta=5/3$ in the longitudinal
   direction (c) and $\sigma=1+2\chi+\zeta=5/4$ in the transverse
   direction (d) ($L=4800$).}
  \label{fig:corre-cst-density}
\end{figure}

The correlation functions obtained numerically are shown in
Fig.~\ref{fig:corre-cst-density}.  All numerical details are given in
\cite{Supp}.  In Fig.~\ref{fig:corre-cst-density}(a), we first compare
results obtained in the isotropic Eq.~(\ref{eq:full-eq}) (with an
external field applied) and in our Eqs.~(\ref{eq:theta2d-finale})
and~(\ref{eq:theta2d-finale-exp}).  We show correlations in the
longitudinal direction with $\qperp=0$ and in the transverse direction
with $\qpar=0$.  We find them nearly identical in all cases except for
Eq.~(\ref{eq:full-eq}) in the parallel direction. We believe this
discrepancy to come from the fluctuations of the global direction from
which, as discussed above, the isotropic equation suffers. In
Fig.~\ref{fig:corre-cst-density}(b-d), we use
Eq.~(\ref{eq:theta2d-finale-exp}) which is the most efficient
numerically. Fig.~\ref{fig:corre-cst-density}(c,d) shows the spatial
correlations for different noise levels, rescaled by the predicted
exponents (for comparison we show the ``raw'' data and the data
rescaled by the exponents of~\cite{toner_birth_2012} in~\cite{Supp}).
The predicted scaling is observed at small enough wave-vector, above a
length scale that seems minimal for $\Delta\approx 4$.~\footnote{The
  large crossover scale at small noise explains why we erroneously
  concluded before~\cite{besse_metastability_2022} that the scaling of
  fluctuations in Malthusian flocks were consistent with the
  predictions of Ref.~\cite{toner_birth_2012}.}.  The dynamic exponent
$z$ is accessed by measuring the space-time correlations
$\langle |\theta(\vec q,\omega)|^2\rangle$.  The width $\Delta \omega$
of the unique propagative mode scales as $\qpar^{z/\zeta}$ and
$\qperp^z$ in the longitudinal and transverse directions respectively
(Fig.~\ref{fig:corre-cst-density}(b)). Fitting the small-$q$ behavior
in Fig.~\ref{fig:corre-cst-density}(b-d), we arrive at the values
reported in Table~\ref{tab:exponents-malth} for the scaling exponents,
in very good agreement with the theoretical predictions.

We now turn to the Vicsek universality class, when the Goldstone mode
$\vec n$ is coupled to a conserved scalar field $c$ representing
density fluctuations. The dynamics of $\vec n$ are still given by
Eqs.~(\ref{eq:rotation}-\ref{eq:all-terms-R}), except that now all
coefficients depend on density in Eq.~(\ref{eq:all-terms-R}). 
In addition, at the same order in gradients and order $2$ in $c$, 
we have~\footnote{We could in principle complement Eq.~(\ref{eq:density}) with a
conserved noise term but it is always irrelevant compared to the
non-conserved noise on $\theta$.}:
\begin{align}
  \partial_t c&=-\div\left[-D_c\nabla c+(v_0+v_1c+v_2 c^2 \right. \nonumber\\ & \qquad\qquad\qquad\left. +w_1\div \vec n+w_2\vec n\cdot\nabla c)\vec n\right]
  \label{eq:density}
\end{align}
Specializing to $d=2$ and neglecting anisotropic diffusion
terms ($w_1$ and $w_2$ in \eqref{eq:density} and $D_{i>1}$ in \eqref{eq:all-terms-R}), we arrive at
\begin{align}
  \label{eq:theta-with-rho}
  \partial_t\theta=\div\left[(\lambda_0+\lambda_1c+\lambda_2 c^2)\nperp +D\nabla\theta\right]+\sqrt{2\Delta}\xi
\end{align}
where the $\lambda_i$ coefficients come from expanding
$\lambda=\lambda_0+\lambda_1c+\lambda_2 c^2$ in
Eq.~(\ref{eq:all-terms-R}). Neglecting the
higher-order terms $v_2 c^2\dperp\theta$ and $\lambda_2 c^2\dpar\theta$, we finally obtain:
\begin{align}
  \label{eq:vicsek-c}
  \partial_t c&+(v_0+v_1c)\dperp\theta+v_1\dpar c+v_2\dpar c^2=D_c\nabla^2 c\\
  \label{eq:vicsek-theta}
  \partial_t\theta&+(\lambda_0+\lambda_1 c)\dpar\theta=\lambda_1\dperp c+\lambda_2\dperp c^2+D\nabla^2\theta + \sqrt{2\Delta}\xi
\end{align}
where all coefficients are constant. As Eq.~(\ref{eq:theta2d-finale})
did for Malthusian flocks, we believe that
Eqs.~(\ref{eq:vicsek-c},\ref{eq:vicsek-theta}) encompass the universal
physics of flocks with conserved density. Again, we show
in~\cite{Supp} that the Goldstone mode obtained by eliminating the
fast variable from the isotropic Toner-Tu equation is {\it not} the
direction $\phi$ of the velocity field, as naively expected. This has
important consequences for the scaling relations since the
nonlinearities in the equation for $\phi$ are not
derivatives~\cite{toner_reanalysis_2012}.

\begin{table}[t!]
  \caption{Scaling exponents for $d=2$ Vicsek flocks predicted by
    Eq.~(\ref{eq:scaling-relations-full}), in the original Toner and
    Tu article~\cite{toner_long-range_1995}, measured in the Vicsek
    model by Mahault et al.~\cite{mahault_quantitative_2019}, and
    measured in our simulations of
    Eqs.~(\ref{eq:density})~(\ref{eq:theta-with-rho}).}
  \centering
  \begin{ruledtabular}
  \begin{tabular}{ccccc}
    \hspace{1cm} & This work & TT 1995 & Mahault {\it et al.} & Numerics \\
    \hline
    $\chi$  & \!\!\!\!\!$-1/3$ & \!\!\!\!\!$-1/5$ & \!\!\!\!\!$-0.31(2)$ & \!\!\!\!\!$-0.34(3)$\\
    $\zeta$  & $1$  &  $3/5$ & $0.95(2)$ & $1.01(4)$\\
    $z$  & $4/3$ & $6/5$ & $1.33(2)$ & $1.30(6)$\\
  \end{tabular}
    \end{ruledtabular}
  \label{tab:exponents}
\end{table}

\begin{figure}[t!]
  \centering
  \includegraphics[width=\columnwidth]{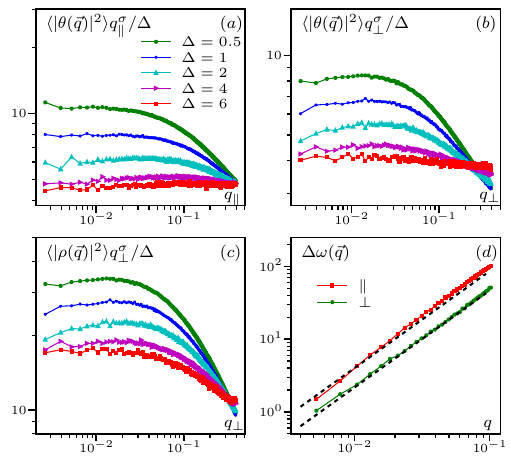}
 \caption{ Correlation of fluctuations in $d=2$ Vicsek flocks from
   numerical integration of
   (\ref{eq:vicsek-c},\ref{eq:vicsek-theta}). (a-c) Static
   correlations rescaled by the exponent expected from our prediction
   $\sigma=4/3$ ($L=4800$). (d) Width of the peaks in the dynamic
   correlation function $\langle |\theta(\vec q,\omega)|^2\rangle$ for
   $\Delta=4$, $L=2400$.  Parameters $v_0=2$, $v_1=1$,
   $\lambda_0=0.5$, $\lambda_1=-0.5$, $\lambda_2=0$, $D_c=D=1$.  }
  \label{fig:corre-density}
\end{figure}

Expanding around the direction of order, assumed to be along
$\vec{e_x}$, and introducing a different coefficient for each term we
obtain at order $2$ in the fields
\begin{align}
  \partial_t c&+v_1\partial_x c+v_0\partial_y\theta+g_x\partial_x\theta^2+g_y\partial_y(\theta c) \nonumber\\
  \label{eq:rhopert}&\qquad +v_2\partial_x c^2=D_{c x}\partial^2_x c+D_{c y}\partial^2_y c\\
  \partial_t\theta&+\lambda_0\partial_x\theta-\lambda_1\partial_y c+h_x\partial_x(\theta c)+h_y\partial_y\theta^2  \nonumber\\
  \label{eq:thetapert} &\qquad -\lambda_2\partial_y c^2=D_x\partial^2_x\theta+D_y\partial^2_y\theta + \sqrt{2\Delta}\xi
\end{align}
with bare values
$g_x=-\frac{v_0}{2}, g_y=v_1, h_x=\lambda_1, h_y=\frac{\lambda_0}{2},
D_{c x}=D_{c y}=D_c, D_x=D_y=D$. Under a RG step with rescaling
\eqref{eq:rescaling}, and $c\to b^\chi c$, the coefficients of
Eqs.~(\ref{eq:rhopert},\ref{eq:thetapert}) flow as
\begin{align}
 &\frac{dg_x}{ds}=[\chi \!+\! z \!-\! \zeta \!+\! \eta_{g_x}] \, g_x \,; \;\;\;
 \frac{dg_y}{ds}=[\chi \!+\! z \!-\! 1\!+\! \eta_{g_y}] \, g_y \,; \nonumber\\
  &\frac{dh_x}{ds}=[\chi \!+\! z \!-\! \zeta \!+\! \eta_{h_x}]\, h_x \,; \;\; 
  \frac{dh_y}{ds}=[\chi \!+\! z \!-\! 1 \!+\! \eta_{h_y}] \, h_y \,; \nonumber\\
  &\frac{d\lambda_2}{ds}=[\chi \!+\! z \!-\! 1 \!+\! \eta_{\lambda_2}] \,\lambda_2 \,;\, \;\;
  \frac{dv_2}{ds}=[\chi \!+\! z \!-\! \zeta \!+\! \eta_{v_2}]\,v_2 \,;\nonumber\\
  &\frac{d\Delta}{ds}=\tfrac{1}{2} [z \!-\! 2\chi \!-\! 1 \!-\! \zeta + 2\eta_{\Delta}] \,\Delta \;.
  \label{eq:RG-flow-full}
\end{align}
The diffusion coefficients flow similarly as in
Eq.~(\ref{eq:RG-flow}). 

Remarkably, as in the Malthusian case, the scaling exponents can be
derived exactly. First, as before, $\eta_\Delta=0$ because all
nonlinearities are derivatives. Furthermore, note that the linear
drift terms with coefficient $v_{0,1}$ and $\lambda_{0,1}$ in
(\ref{eq:rhopert},\ref{eq:thetapert}) are of lower order than those
included in Eq.~(\ref{eq:RG-flow-full}), and are thus diverging at a
fixed point. These coefficients control the mode structure which
contains two normal modes $\psi_+$ and $\psi_-$ mixing $\theta$ and
$c$~\cite{toner_flocks_1998,geyer_sounds_2018}. As detailed
in~\cite{Supp}, because the propagators associated to $\psi_\pm$ peak
at different frequencies with vanishingly small overlap at small
wavelength, their dynamics essentially decouple. They then separately
obey Galilean invariance, forbidding the renormalization of the
quadratic nonlinearities $g_{x,y}$, $h_{x,y}$, $v_2$ and $\lambda_2$.
In the end, none of the terms in Eq.~(\ref{eq:RG-flow-full}) receive
graphical corrections so that, at the infrared fixed point, one has
the scaling relations
\begin{equation}
  \label{eq:scaling-relations-full}
  \chi+z-\zeta=\chi+z-1=z-2\chi-1-\zeta=0
\end{equation}
yielding the exponents reported in Table~\ref{tab:exponents}.

The predicted values are found to be in agreement with the exponents
measured in the Vicsek model~\cite{mahault_quantitative_2019} also
reported in Table \ref{tab:exponents}. In addition, we performed
numerical simulations of (\ref{eq:vicsek-c},\ref{eq:vicsek-theta}) for
an arbitrarily chosen set of parameters varying the noise intensity
$\Delta$. We show in Fig.~\ref{fig:corre-density} the same observables
as in Fig.~\ref{fig:corre-cst-density} with the addition of the
density correlations $\langle |c(q_\parallel=0,q_\bot,t)|^2\rangle$ in
the transverse direction, expected to scale like the correlations of
$\theta$. In the longitudinal direction, the correlations of $c$ have
a more complex scaling. It will be the topic of future work to examine
these, as well as exploring more broadly the parameter space of
Eqs.~(\ref{eq:vicsek-c},\ref{eq:vicsek-theta}). The trend seen in
Fig.~\ref{fig:corre-density} is the same as for Malthusian flocks: to
a good approximation the predicted scaling is reached at large
distance after a small-scale regime with a crossover length that his
minimal for noise intensity $\Delta\approx 6$. Extracting numerical
values by fitting the large scale behavior as detailed in~\cite{Supp}
give the numbers reported in Table~\ref{tab:exponents}, consistent
with the predictions.

{\it Conclusion.} We have proposed a hydrodynamic theory for the
ordered phase of polar flocks based on writing generic equations
compatible with the symmetry of these phases. We focused in particular
on the $d=2$ case without density field (Malthusian class) and with
density (Vicsek class).  In addition to exhibiting clearly the
symmetries, our equations provide an efficient numerical platform to
measure correlations of the fluctuations in the ordered phase.
Importantly, we have derived exact scaling exponents for the two
universality classes in $d=2$ and found good agreement with the
numerical results.

During the writing of this manuscript, two directly relevant preprints
have appeared. In the first one~\cite{ikeda_how_2024}, Ikeda obtains
in the Malthusian case the exponents of~\cite{toner_birth_2012}
reported in Table~\ref{tab:exponents-malth}, which we have shown to
fail, but obtains the same exponent as we do in the Vicsek case,
although from different arguments. In~\cite{jentsch_new_2024}, Jentsch
and Lee tackle the Vicsek case with the non-perturbative
renormalization group~\cite{delamotte_introduction_2012}. They are
able to derive exact scaling exponents in arbitrary dimension,
different from ours in $d=2$ (although also compatible with the
numerics of~\cite{mahault_quantitative_2019}), but at the price of
neglecting several nonlinearities ($g_x$, $g_y$, $v_2$, $h_x$ and
$\lambda_2$ in Eq.~(\ref{eq:thetapert}). However, these nonlinearities
being relevant in the RG sense, we have all reasons to believe that
they would break the scaling relations derived
in~\cite{jentsch_new_2024}.

We believe that our approach opens new possibilities to investigate
the symmetry-broken phase of active systems. Future work will
be devoted to investigate higher dimensions and other symmetry classes
including incompressible flocks and active nematics.

\acknowledgements We thank B. Delamotte, J. Horowitz, P. Jentsch,
C. F. Lee, B. Mahault, M. Tissier and N. Wschebor for insightful
discussions.

\bibliography{refs.bib}


\section*{Supplementary material}

# Supplementary information for "Dynamic Scaling of Two-Dimensional Polar Flocks"

Hugues Chaté[1, 2, 3] and Alexandre Solon[3]

[1]*Service de Physique de l'Etat Condensé, CEA, CNRS Université Paris-Saclay, CEA-Saclay, 91191 Gif-sur-Yvette, France*
[2]*Computational Science Research Center, Beijing 100094, China*
[3]*Sorbonne Université, CNRS, Laboratoire de Physique Théorique de la Matière Condensée, 75005 Paris, France*
(Dated: March 6, 2024)

## I. NUMERICAL DETAILS

The equations are integrated using a semi-spectral algorithm with explicit Euler time-stepping. We use large values of spatial resolution up to $dx = 8$ to speed up the simulations since, as shown in Fig. S1(a,b), this affects only the short-distance physics, leaving the large-scale behavior unchanged.

We show in Fig. S1(c,d) the finite size scaling of the static correlations for a relatively low noise level $\Delta = 1$ for which, on accessible system sizes, the system is not in the scaling regime. As system size increases, we see that the correlations slowly converge toward the expected asymptotic scaling. To reach the asymptotic regime, one would need still much larger system sizes or, alternatively, to tune the noise level in order to minimize the crossover length, as we do in the main text Fig. 1 (c-d) and Fig. 2 (a-c).

Parameters of main text figures:

- Fig. 1 (a): $\Delta = 0.5$, $\lambda = D = 1$, $dx = 4$ and $dt = 0.2$ for all curves. System size $L = 2400$ for the isotropic equation, $L = 4800$ otherwise. As explained in the main text, to compute anisotropic correlation functions, we need to fix the direction of global order along one axis of the simulation box. For the isotropic equation, we thus apply an external field of strength $h = 3 \times 10^{-6}$ along the $x$-axis.

- Fig. 1 (b): $L = 2400$, $\lambda = D = 1$, $\Delta = 4$, $dx = 8$, $dt = 0.1$.

- Fig. 1 (c-d): $L = 4800$, $\lambda = D = 1$, $dx = 8$, $dt = 0.2/\Delta$.

- Fig. 2: $v_0 = 2$, $v_1 = 1$, $\lambda_0 = 0.5$, $\lambda_1 = -0.5$, $\lambda_2 = 0$, $D_c = D = 1$, $dx = 8$, $dt = 0.05/\Delta$. $L = 4800$ (a-c) or $L = 2400$ (d) and $\Delta = 2$ (d).

## II. SCALING EXPONENTS

### A. Numerical evaluation

To evaluate numerically the scaling exponents that are reported in Tables I and II (main text), we fit the small $q$ behavior of $\langle|\theta(\mathbf{q},t)|^2\rangle$ in the longitudinal and transverse directions and the width of the peaks of dynamical correlations $\Delta\omega$ in the transverse direction. The fits are done on the data with the noise level $\Delta = 4$ that has a short crossover length. Repeating the fits on the curve with lower noise level gives us an estimate of the error bars. For

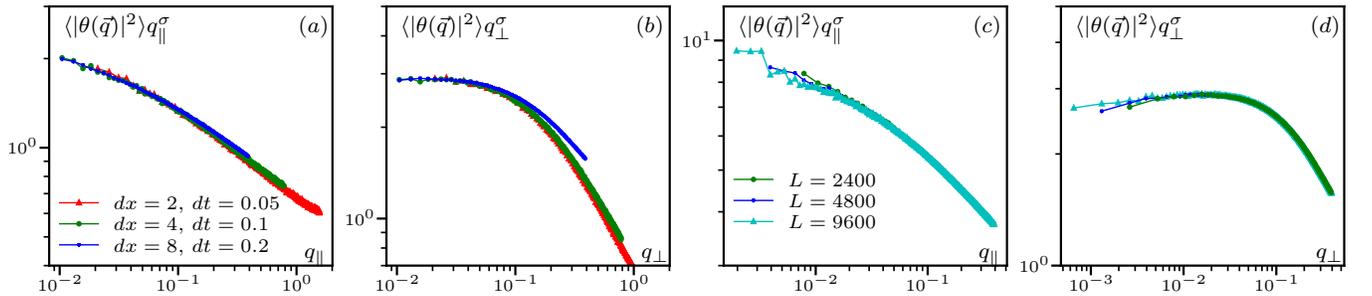

FIG. S1. (a,b): Static correlation functions for varying numerical accuracy $dx$ and $dt$, rescaled by the predicted scaling exponent. $L = 1200$ for $dx = 2$ and $L = 2400$ for $dx = 4, 8$. (c,d): Static correlations for varying system size. $dx = 8$, $dt = 0.2$. All panels: simulation of Eq. (8) (main text) with $\lambda = D = \Delta = 1$.

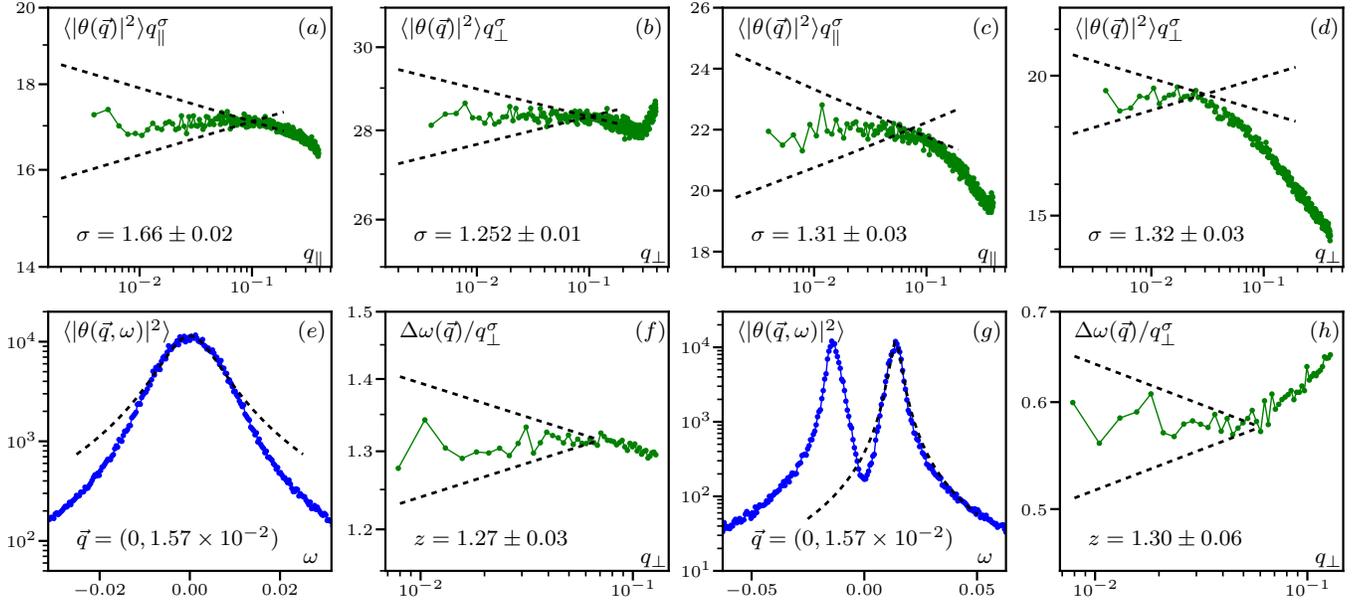

FIG. S2. Numerical evaluation of scaling exponents. Data rescaled by the fitted value of the exponent, indicated on the figure. The black dashed lines materialize the error bars. (a-d): Static correlations in the Malthusian (a,b) and Vicsek (c,d) case. (e,g): Dynamic correlation function from which the width $\Delta\omega$ is extracted by fitting $\langle|\theta(\mathbf{q},\omega)|^2\rangle = A/(1+(\omega/\Delta\omega)^2)$ in the Malthusian (e) and Vicsek (g) case. (f,h): Scaling of $\Delta\omega$ in the Malthusian case (f) and Vicsek (h) case.

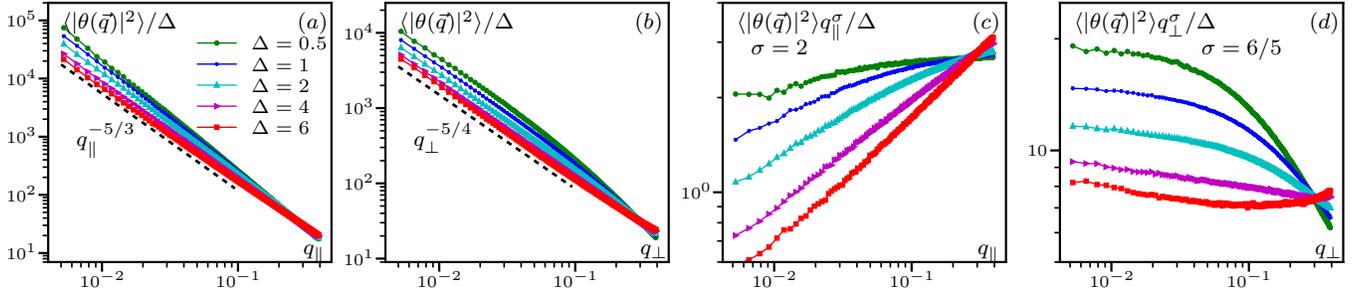

FIG. S3. Static correlations functions in $d=2$ Malthusian flocks. Same data as in Fig. 1 (c,d) (main text). (a,b): Comparison with the predicted scaling (black dashed line). (c,d): Data rescaled by the exponents predicted in [1].

each fit, we show the data rescaled by the fitted exponent with the error bars in Fig. S2. We see that the estimations of the error bars are consistent with the precision of the data.

### B. Non-renormalization arguments

#### 1. Malthusian case

We find that Eq. (8) (main text) is invariant under the transformation

$$\theta'(\mathbf{r},t) = \theta(\mathbf{r}-\mathbf{r}_c, t) + \theta_0; \qquad \partial_t \mathbf{r_c} = \theta_0 \begin{pmatrix} 2\lambda_x \theta \\ \lambda_y \end{pmatrix} \qquad (1)$$

for arbitrary transformation parameter $\theta_0$. When $\lambda_x = 0$, Eq. (1) is a standard Galilean transformation which must be satisfied with the same value of $\lambda_y$ along the RG flow and thus impose that $\lambda_y$ does not receive graphical corrections, like in the Burgers [2], Navier-Stokes [3] or KPZ [4] equations.

We expect that the generalized Galilean invariance Eq. (1) lead to the same type of constraint, requiring that both $\lambda_x$ and $\lambda_y$ are not graphically corrected. Writing the associated Ward-Takahashi identities is, however, a non-trivial

task since Eq. (1) is a non-linear symmetry, that we defer to future work.

### 2. Vicsek case

The mode structure generated by the linearized hydrodynamics is well known [5, 6]. It features two propagating sound modes that can be computed from the linearized Eq. (17,18) (main text) in Fourier space and time, for frequency $\omega$ and wavevector $\mathbf{q} = (q_x, q_y)$

$$\mathbb{M} \cdot \begin{pmatrix} c(\mathbf{q},\omega) \\ \theta(\mathbf{q},\omega) \end{pmatrix} = 0; \qquad \mathbb{M} = \begin{pmatrix} i(\omega - q_x v_1) + D_{cx} q_x^2 + D_{cy} q_y^2 & -i q_y v_0 \\ i q_y \lambda_1 & i(\omega - q_x \lambda_0) + D_x q_x^2 + D_y q_y^2 \end{pmatrix} \qquad (2)$$

The eigenvalues of $\mathbb{M}$ give two sound modes. At small wavevector $\mathbf{q} = q(\cos\alpha, \sin\alpha)$, their frequency can be written

$$\omega_\pm = \left(c_\pm(\alpha) q + O(q^3)\right) + i\Delta\omega_\pm \qquad (3)$$

with $\Delta\omega_\pm = O(q^2)$ associated with the damping of the wave and the velocities

$$c_\pm(\alpha) = \frac{v_1 + \lambda_0}{2} \pm \frac{1}{2}\sqrt{(v_1 - \lambda_0)^2 \cos^2\alpha - 4 v_0 \lambda_1 \sin^2\alpha}. \qquad (4)$$

The associated eigenmodes $\psi_\pm$ are linear combinations mixing $c$ and $\theta$. Most importantly, the associated propagators $G_0^\pm = (i(\omega - \omega_\pm(\mathbf{q})))^{-1}$ are peaked around the frequencies $\omega_\pm$ with an overlap between the two modes that becomes negligible at small wavenumber since the width of the peaks grows as $q_y^z$ with $z > 1$ [5]. By the same reasoning as explained in [5] (concerning longitudinal and transverse velocities), the normal modes $\psi_+$ and $\psi_-$ essentially decouple because all terms mixing $\psi_+$ and $\psi_-$ in their dynamical equation include products of propagators and correlation functions of the two modes that have vanishingly small overlap.

Writing only the lowest-order terms, in Fourier space, the dynamics of $\psi_\pm$ then read

$$\psi_\pm(\mathbf{q},\omega) = G_0^\pm(\mathbf{q},\omega) \left[ \eta_\pm(\mathbf{q},\omega) + i(a_\pm q_x + b_\pm q_y) \int d\mathbf{q}' d\omega' \psi_\pm(\mathbf{q}',\omega') \psi_\pm(\mathbf{q}-\mathbf{q}',\omega-\omega') \right] \qquad (5)$$

with $G_0^\pm = (i(\omega - \omega_\pm(\mathbf{q})))^{-1}$ and coefficients $a_\pm(\alpha)$ and $b_\pm(\alpha)$ which are combinations of the coefficients of Eq. (17,18) (main text) that can depend on the direction $\alpha$ of the wave-vector but not its amplitude since it would lead to higher order gradient terms. From Eq. (5), one can see that $\psi_+$ and $\psi_-$ are separately invariant under the Galilean transformation which reads in Fourier space

$$\mathbf{q}' = \mathbf{q}; \qquad \omega' = \omega - \mathbf{v} \cdot \mathbf{q}; \qquad \psi'_\pm(\mathbf{q}',\omega') = \psi_\pm(\mathbf{q},\omega) + \varepsilon \delta^2(\mathbf{q})\delta(\omega); \qquad \mathbf{v} = 2 \begin{pmatrix} a_\pm \\ b_\pm \end{pmatrix} \varepsilon \qquad (6)$$

for any transformation parameter $\varepsilon$. Again, the Galilean invariance prevents the renormalization of the coefficients $a_\pm$ and $b_\pm$ and thus of the coefficients $g_{x,y}$, $h_{x,y}$, $v_2$ and $\lambda_2$ in Eq. (17,18) (main text).

## III. ELIMINATION OF THE FAST DEGREE OF FREEDOM

### A. Malthusian flocks

In this section we show how to recover our generic hydrodynamic Eq. (6) (main text) starting from the isotropic "Toner-Tu" equation and integrating the fast degrees of freedom. Expanding Eq. (6) (main text) and putting $D_6 = D_7 = 0$ without loss of generality (they play the same role as $D_3$ and $D_2$ respectively in $d = 2$), it reads

$$\partial_t \theta + \lambda \partial_\parallel \theta = D_1 \nabla^2 \theta + D_3 \partial_\parallel \partial_\parallel \theta - D_2 \partial_\perp \partial_\perp \theta + (D_2 + D_3) \partial_\parallel \theta \partial_\perp \theta + \sqrt{2\Delta}\xi \qquad (7)$$

with Gaussian white noise of unit variance $\xi$. As shown in the main text, this is the most general equation that should be obeyed by the phase of a Goldstone mode in $d = 2$. We thus expect to recover the same equation starting from the isotropic equation and eliminating the fast degree of freedom.

Let us hence consider the most general equation for a vector field that obeys the rotational $O(d = 2)$ symmetry. We keep terms up to order 2 in gradients and order 3 in the field $\mathbf{v}$. We thus have a more general version of Eq. (1) (main text)

$$\partial_t \mathbf{v} + \lambda_1 (\mathbf{v} \cdot \nabla)\mathbf{v} + \lambda_2 (\nabla \cdot \mathbf{v})\mathbf{v} + \lambda_3 \nabla(|\mathbf{v}|^2) = D\nabla^2 \mathbf{v} + D_B \nabla(\nabla \cdot \mathbf{v}) + \alpha(v_0^2 - |\mathbf{v}|^2)\mathbf{v} + \sqrt{2\Delta}\,\boldsymbol{\xi}, \qquad (8)$$

with Gaussian white noise of unit variance $\boldsymbol{\xi}$ as in the main text. Let us parameterize the velocity vector as $\mathbf{v} = v(\mathbf{r},t)\mathbf{n}(\phi(\mathbf{r},t))$ with $\mathbf{n}(\phi) = (\cos\phi, \sin\phi)$ the unit vector pointing in direction $\phi$. We can rewrite Eq. (8) for the two variables $v$ and $\phi$, using that $\partial_t \mathbf{v} = \partial_t v \mathbf{n} + v \partial_t \phi \mathbf{n}_\perp$ with $\mathbf{n}_\perp = (-\sin\phi, \cos\phi)$, projecting on $\mathbf{n}$ and $\mathbf{n}_\perp$ gives

$$\partial_t v + \lambda_1 v \partial_\parallel v + \lambda_2(v\partial_\parallel v + v^2 \partial_\perp \phi) + 2\lambda_3 v \partial_\parallel v = D\left(\nabla^2 v - v(\nabla\phi)^2\right) + D_B \partial_\parallel \left(v\partial_\perp \phi + \partial_\parallel v\right) + \alpha(v_0^2 - v^2)v + \sqrt{2\Delta}\xi_\parallel \tag{9}$$

$$\partial_t \phi + \lambda_1 v \partial_\parallel \phi + 2\lambda_3 \partial_\perp v = D\left(\nabla^2 \phi + \frac{2\nabla v \cdot \nabla \phi}{v}\right) + \frac{D_B}{v}\partial_\perp \left(v\partial_\perp \phi + \partial_\parallel v\right) + \sqrt{2\Delta}\xi_\perp / v \tag{10}$$

where $\partial_\parallel = \mathbf{n}_\parallel \cdot \nabla$, $\partial_\perp = \mathbf{n}_\perp \cdot \nabla$ and $\xi_\parallel$ and $\xi_\perp$ are the two components of $\boldsymbol{\xi}$ in Eq. (8).

The norm of the velocity is a fast variable which relaxes in a time of order $(\alpha v_0^2)^{-1}$. We can eliminate it, formally by doing an expansion at large $\alpha$. We want to retain terms only up to order $\nabla^2$ in the equation for $\phi$. We thus keep the terms up to order $\nabla$ for $v$,

$$v \approx v_0 - \frac{\lambda_2}{2\alpha}\partial_\perp \phi. \tag{11}$$

Inserting in Eq. (10), we obtain

$$\partial_t \phi + \lambda \partial_\parallel \phi = D\nabla^2 \phi + D_\perp \partial_\perp \partial_\perp \phi + D_\times \partial_\parallel \phi \partial_\perp \phi + \sqrt{2\tilde{\Delta}}\xi_\perp \tag{12}$$

with $\lambda = v_0 \lambda_1$, $D_\perp = D_B + \lambda_2 \lambda_3/\alpha$, $D_\times = \lambda_1 \lambda_2/(2\alpha)$ and $\tilde{\Delta} = \Delta/v_0^2$.

Interestingly, Eq. (12) does not have exactly the same form as Eq. (7) because the coefficients of the diffusion terms have a different structure: the terms $D_\perp$ and $D_\times$ differ from the three diffusion terms with coefficient $D_{2,3}$ in Eq. (7). However, we can easily recover the structure of Eq. (7) by considering the combination $\tilde{\phi} = \phi - \alpha \partial_\parallel \phi$. At the same order in gradient, it follows that

$$\partial_t \tilde{\phi} + \lambda \partial_\parallel \tilde{\phi} = D\nabla^2 \tilde{\phi} + D_\perp \partial_\perp \partial_\perp \tilde{\phi} + (D_\times - \alpha\lambda)\partial_\parallel \tilde{\phi}\partial_\perp \tilde{\phi} + \sqrt{2\tilde{\Delta}}\xi_\perp \tag{13}$$

Choosing $\alpha = (D_\perp + D_\times)/\lambda$, Eq. (13) is exactly Eq. (7) upon identifying $D_1 = D$, $D_2 = -D_\perp$ and $D_3 = 0$. The fact that we had to consider a different variable $\tilde{\phi}$ to make the identification suggests that the phase $\phi$ of the velocity vector is not the correct Goldstone mode to consider in the ordered phase. Note that this change of variable makes sense only for an active system in which the direction of order couples to spatial directions.

### B. Compressible flocks

Let us repeat the same procedure in the case when the velocity is coupled to the density field. We consider the Toner-Tu equations

$$\partial_t \mathbf{v} + \lambda_1 (\mathbf{v}\cdot\nabla)\mathbf{v} + \lambda_2 (\nabla\cdot\mathbf{v})\mathbf{v} + \lambda_3 \nabla(|\mathbf{v}|^2) = -\nabla P(\rho) + D\nabla^2 \mathbf{v} + D_B \nabla(\nabla\cdot\mathbf{v}) + \alpha(v_0^2 - |v|^2)\mathbf{v} + \sqrt{2\Delta}\,\boldsymbol{\xi}$$
$$\partial_t \rho + \nabla\cdot(\rho\mathbf{v}) = 0 \tag{14}$$

with pressure $P(\rho) = \sigma_1 \rho + \sigma_2 \rho^2$. As before we parameterize the velocity vector as $\mathbf{v} = v(\mathbf{r},t)\mathbf{n}(\phi(\mathbf{r},t))$. We also expand around the homogeneous density $\rho(\mathbf{r},t) = \rho_0 + c(\mathbf{r},t)$. In principle all coefficients in Eq. (14) depend on density. For simplicity, we retain only that of $\lambda_1 = \lambda_{10} + \lambda_{1c}c$, which is enough to generate the important nonlinearities [7]. We follow the same procedure as in the previous section to eliminate the norm of the velocity vector. Solving the equation for $v$ at order $O(v,c,\nabla)$ now gives

$$v \approx v_0 - \frac{\lambda_2}{2\alpha}\partial_\perp \phi - \frac{\sigma}{2v_0^2 \alpha}\partial_\parallel c \tag{15}$$

with $\sigma = \sigma_1 + 2\rho_0\sigma_2$. The remaining equations for $\phi$ and $c$ then read

$$\partial_t \phi + \lambda \partial_\parallel \phi + gc\partial_\parallel \phi + \tilde{\sigma}\partial_\perp c = -\tilde{\sigma}_2 \partial_\perp c^2 + D\nabla^2 \phi + D_\perp \partial_\perp \partial_\perp \phi + D_\times \partial_\parallel \phi \partial_\perp \phi$$
$$+ D_{\phi c}(\partial_\parallel \phi \partial_\parallel c - \partial_\perp \phi \partial_\perp c) + D_{cc}\partial_\perp c \partial_\parallel c + \sqrt{2\tilde{\Delta}}\xi_\perp \tag{16}$$

$$\partial_t c + \nabla\cdot\left[(\rho_0 v_0 + v_0 c + w_1 \partial_\perp \phi + w_2 \partial_\parallel c)\mathbf{n}\right] = 0 \tag{17}$$

with coefficients

$$\lambda = v_0 \lambda_{10}; \quad g = v_0 \lambda_{1c}; \quad \tilde{\sigma} = \frac{\sigma}{v_0}; \quad \tilde{\sigma}_2 = \frac{\sigma_2}{v_0}; \quad D_\perp = D_B + \frac{\lambda_2 \lambda_3}{\alpha}; \quad D_\times = \frac{\lambda_{10} \lambda_2}{2\alpha}; \quad (18)$$

$$D_{\phi c} = \frac{\lambda_{10} \sigma}{2 v_0^2 \alpha}; \quad D_{cc} = -\frac{\sigma^2}{2 v_0^4 \alpha}; \quad \tilde{\Delta} = \frac{\Delta}{v_0^2}; \quad w_1 = -\frac{\lambda_2 \rho_0}{2\alpha}; \quad w_2 = -\frac{\sigma \rho_0}{2 v_0^2 \alpha}; \quad (19)$$

We want to compare with the generic hydrodynamic equations for a Goldstone mode coupled to a conserved density field, as constructed in the main text. Compared to the Malthusian case, all parameters can depend on density. Starting from Eq. (6) (main text) (again with $D_6 = D_7 = 0$ without loss of generality) we consider coefficients $\lambda = \lambda^{(0)} + \lambda^{(1)} c + \lambda^{(2)} c^2 + \lambda^{(3)} \partial_\parallel c$, $D_i = D_i^{(0)} + D_i^{(1)} c$ with $i = 1, 2, 3$ (we depart from the notation of the main text to avoid confusion with the coefficients of the Toner-Tu equation). We obtain the analog of Eq. (7) when density is taken into account

$$\partial_t \theta + \lambda^{(0)} \partial_\parallel \theta - \lambda^{(1)} \partial_\perp c + \lambda^{(1)} c \partial_\parallel \theta = \lambda^{(2)} \partial_\perp c^2 + D_1^{(0)} \nabla^2 \theta + D_3^{(0)} \partial_\parallel \partial_\parallel \theta - D_2^{(0)} \partial_\perp \partial_\perp \theta + (D_2^{(0)} + D_3^{(0)}) \partial_\parallel \theta \partial_\perp \theta \quad (20)$$

$$+ D_1^{(1)} \nabla c \cdot \nabla \theta + D_3^{(1)} \partial_\parallel c \partial_\parallel \theta + D_2^{(1)} \partial_\perp c \partial_\perp \theta + \lambda^{(3)} \partial_\parallel c \partial_\perp c + \sqrt{2\Delta} \xi \quad (21)$$

As in the Malthusian case, we find that Eq. (16) obtained from the Toner-Tu equation differs from the expected hydrodynamic Eq. (20). They differ, as before, in the structure of the diffusion terms $D_2^{(0)}$ and $D_3^{(0)}$ but also, more crucially, in the structure of the $\lambda^{(1)}$ terms. Indeed, the terms $\partial_\perp c$ and $c\partial_\parallel \theta$ have the same coefficient $\lambda^{(1)}$ in Eq. (20) (because of the underlying gradient structure) while the two terms have different coefficients in Eq. (16). This again indicates that $\phi$ is not a Goldstone mode and thus not the correct variable to consider in a hydrodynamic theory. One should instead consider

$$\tilde{\phi} = \phi - \alpha \partial_\parallel \phi - \beta c \partial_\parallel \phi; \qquad \alpha = \frac{D_\perp + D_\times}{\lambda}; \qquad \beta = \frac{g - \tilde{\sigma}}{\lambda} \quad (22)$$

which dynamics follows Eq. (20) and is thus a Goldstone mode. Note that there is no problem for the $c$ equation since the same terms are found in both Eq. (17) and in the hydrodynamic theory Eq. (15) (main text), as expected since $c$ is obviously a conserved quantity and thus a proper slow mode on which to build a hydrodynamic theory.

### C. Hydrodynamic theory with the naive choice of variables

For concreteness, let us explore what happens when taking Eq. (16,17) as our hydrodynamic theory, as is done *e.g.* in [7]. Expanding around $\phi = 0$ in powers of $\phi$ leads, compared to Eq. (17,18) (main text), to more general terms $h_{x,1} c \partial_x \phi + h_{x,2} \phi \partial_x c$ with $h_{x,1} \neq h_{x,2}$ in general, instead of the derivative term $h_x \partial_x (c\theta)$. We simulated this equation with the same parameters as in Fig. 2 (main text) and varying $h_{x,2}$. As shown in Fig. S4, we find that only for $h_{x,2} = h_{x,1}$ do we observe the expected scale-free fluctuations. For $h_{x,2} < h_{x,1}$ we find that the equation is numerically unstable while for $h_{x,2} > h_{x,1}$, the system shows short-ranged correlations, signaled by the small-$q$ plateau in the correlation functions. The plateau is reached beyond a crossover length that diverges as $h_{x,2} \to h_{x,1}$. These results are consistent with the fact that $\phi$ is not a slow variable, except in the special case $h_{x,2} = h_{x,1}$.

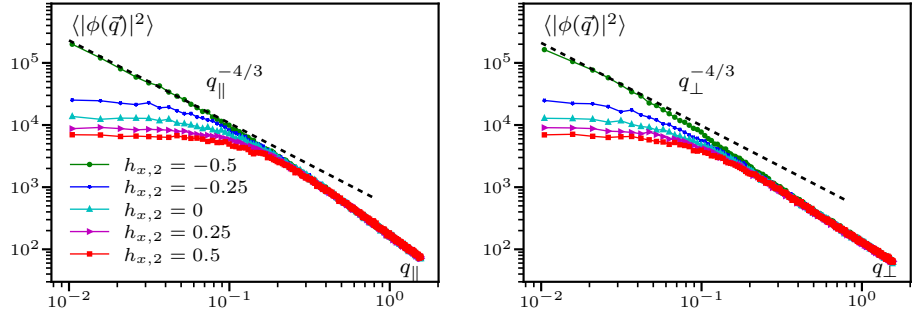

FIG. S4. Correlation of fluctuations from numerical integration of the equations for $c$ and $\phi$ with a non-derivative term $h_{x,1} c\partial_x \phi + h_{x,2} \phi \partial_x c$ in the dynamics of $\phi$, as described in Sec. III C. We use the same parameter values as in Fig. 2 (main text) $v_0 = 2$, $v_1 = 1$, $\lambda_0 = 0.5$, $\lambda_1 = -0.5$, $\lambda_2 = 0$, $D_c = D = 1$ with $h_{x,1} = \lambda_1 = -0.5$. For all $h_{x,2} < h_{x,1}$, we have found that the system is numerically unstable. $L = 1200$, $dx = 2$, $dt = 0.05$.



\end{document}